# Did Hurricane Katrina Reduce Mortality?

By Robert Kaestner

**Abstract**

In a recent article in the American Economic Review, Tatyana Deryugina and David Molitor (DM) analyzed the effect of Hurricane Katrina on the mortality of elderly and disabled residents of New Orleans. The authors concluded that Hurricane Katrina improved the eight-year survival rate of elderly and disabled residents of New Orleans by 3% and that most of this decline in mortality was due to declines in mortality among those who moved to places with lower mortality. In this article, I provide a critical assessment of the evidence provided by DM to support their conclusions. There are three main problems. First, DM generally fail to account for the fact that people of different ages, races or sex will have different probabilities of dying as time goes by, and when they do allow for this, results change markedly. Second, DM do not account for the fact that residents in New Orleans are likely to be selected non-randomly on the basis of health because of the relatively high mortality rate in New Orleans compared to the rest of the country. Third, there is considerable evidence that among those who moved from New Orleans, the destination chosen was non-random. Finally, DM never directly assessed changes in mortality of those who moved, or stayed, in New Orleans before and after Hurricane Katrina. These problems lead me to conclude that the evidence presented by DM does not support their inferences.

In a recent article in the *American Economic Review*, Tatyana Deryugina and David Molitor (2020) analyzed the effect of Hurricane Katrina, which occurred in New Orleans in late August 2005, on mortality of residents of New Orleans enrolled in Medicare. The article is titled, "Does When You Die Depend on Where You Live? Evidence from Hurricane Katrina."

I reproduce the abstract in full and apply boldface to parts:

We follow Medicare cohorts to estimate Hurricane Katrina's long-run mortality effects on victims initially living in New Orleans. Including the initial shock, the hurricane improved eight-year survival by 2.07 percentage points. **Migration to lower-mortality regions explains most of this survival increase.** Those migrating to low-versus high-mortality regions look similar at baseline, but their subsequent mortality is 0.83–1.01 percentage points lower per percentage point reduction in local mortality, quantifying **causal effects** of place on mortality among this population. Migrants' mortality is also lower in destinations with healthier behaviors and higher incomes but is unrelated to local medical spending and quality. (Deryugina and Molitor 2020, 3602, boldface added)

I offer several criticisms of Deryugina and Molitor (DM), but let me start with two illustrations that highlight the intuition of some of my criticisms.

First, consider a time just prior to Hurricane Katrina, say July 2005, and two men, both age 70, one living in New Orleans and one living in Richmond, Virginia, which is one of DM's comparison regions. Assuming a world like ours except that Hurricane Katrina does *not* happen, do these two men have equal chances of surviving to 2013? The answer may well be no: The New Orleans man is someone who has made it to 70 *in New Orleans*, which has a relatively high mortality rate. Making it to 70 there suggests unusually good health compared to someone who made it to age 70 in Richmond, which has a relatively low mortality rate. Thus, it is likely that the 70-year-old in New Orleans will live longer than the 70-year-old in Richmond. If so, then an analysis of the effect of Hurricane Katrina will appear to lower death rates when in fact it is simply an artifact of not comparing people with the same health.

Now consider a second scenario that involves two 70-year-olds in New Orleans in July 2005. After Hurricane Katrina, one moves to Baton

Rouge because a city bus was available to take them there, while the other moves to Houston because he had a car and went to see relatives. In 2013 the man who moved to Baton Rouge has passed away, while the man who moved to Houston still lives. It is possible that the difference in their surviving was in part causally related to the conditions and environments of Baton Rouge versus Houston. But it is also possible that the causal difference was a difference in the two men, a difference that figured into the one's moving to Baton Rouge and the other's moving to Houston. If entirely so, then the man who moved to Baton Rouge and died would have died even if he had moved to Houston.

These two illustrations highlight that people differ in ways that scientists do not and often cannot observe and measure. Those differences might relate to things that are observed, however, such as where one has been living, whether one chooses to move, and to where one moves. We are creative beings, each with "a principle of motion of its own" (Smith TMS, 234.17). That principle of motion, that will, moves on the basis of the soul's particularistic opportunities and constraints. Economists should always figure that inscrutable hidden factors lie behind the events that result from human will.

**Who Did or Did Not Die? Mortality of Movers and Stayers**

DM conducted a difference-in-differences analysis of annual mortality that compared the mortality rate of elderly (roughly 80%) and non-elderly, disabled (roughly 20%) Medicare enrollees living in New Orleans in 2004 in years prior to and after Hurricane Katrina to the mortality rate of similar residents in other areas.[1] Based on the results of this analysis, DM conclude that Hurricane Katrina caused a decline in Medicare-enrollee mortality. I shall cast some doubt on this basic claim.

But first, let's grant that Hurricane Katrina caused a decline in Medicare-enrollee mortality. DM say that such a decline came mainly because of a decline in mortality among movers. However, DM never actually analyzed the causal effect of Katrina on the mortality of movers or stayers pre-to-post Katrina. The difference-in-difference approach of DM cannot identify the effect of Hurricane Katrina on those who moved and those who stayed in New Orleans, but only the effect of Hurricane Katrina on the two groups combined. So, the conclusion of DM, "Migration to lower-mortality regions explains most of this survival increase" (p. 3602), is not supported by direct evidence.

---

[1] In most analyses, DM used residents in 10 cities that closely match New Orleans with respect to median earnings, population growth, and racial composition as the comparison group. The 10 cities are: Baltimore, MD; Birmingham, AL; Detroit, MI; Gary, IN; Jackson, MS; Memphis, TN; Newark, NJ; Portsmouth, VA; Richmond, VA; and St. Louis, MO. Other comparison areas are used instead of these 10 cities and estimates using these alternatives are similar. In other analyses, DM used a larger group of areas defined by county or commuting zone.

The fact that the DM analysis is of *all* residents of New Orleans (and other cities) and not just movers or just stayers is important because it seems reasonable to assume that Hurricane Katrina *increased* subsequent mortality of New Orleans stayers. There was widespread damage to infrastructure, including the destruction of most healthcare facilities (e.g., Memorial Medical Center), and a large share of medical professionals were displaced leaving a severe shortage (Rudowitz et al. 2006). DM allude to possible improvements in the healthcare infrastructure in New Orleans subsequent to Hurricane Katrina (see footnote 21 of paper), but several studies suggest that such was not the case. At best, there was marginal improvement in access provided by clinics, but this did not occur for at least a few years afterward, when Federal assistance was provided (Rittenhouse et al. 2012; Cole et al. 2015; Hamel et al. 2015). The Memorial Medical Center (i.e., Charity Hospital), which had served most low-income persons in New Orleans, was replaced by a new hospital, but that did not open until 2015. Also, there was a large decline in employment (and income) in the short run aftermath of Katrina, although by 2007-08 labor income among those who filed tax returns and could be later observed in tax data recovered to pre-Katrina levels (Deryugina et al. 2018). The DM analysis, however, is focused on elderly and disabled who mainly do not work and who suffered from the destruction and displacement caused by Hurricane Katrina. The trauma, along with a devastated health care system, a destroyed infrastructure and a severely bruised economy, all suggest that the mortality rate of elderly/disabled stayers likely increased.

It is possible to obtain an estimate of the effect of Hurricane Katrina on the mortality rate of each group with the help of a few assumptions. First, note that prior to Hurricane Katrina, for example in 2004, the mortality rate of the elderly and disabled in New Orleans was algebraically equal to the following:

$$(1) \quad M_{pre} = \alpha M_{pre}^{stayers} + (1-\alpha) M_{pre}^{movers}$$

Equation (1) simply states that the mortality rate (*M*) in New Orleans prior to Hurricane Katrina (pre) is equal to the share of New Orleans residents who stayed in New Orleans after Katrina (α) multiplied by their mortality rate at that time (i.e., 2004) plus the share of New Orleans residents who moved out of New Orleans after Katrina (1-α) multiplied by their mortality rate at that time. Similarly, for the post-Katrina period (2006 to 2013), the mortality rate of the elderly and disabled in New Orleans was:

$$(2) \quad M_{post} = \alpha M_{post}^{stayers} + (1-\alpha) M_{post}^{movers}$$

The difference in the mortality rate of New Orleans residents is:

$$(3) \quad M_{post} - M_{pre} = \alpha (M_{post}^{stayers} - M_{pre}^{stayers}) + (1-\alpha)(M_{post}^{movers} - M_{pre}^{movers})$$

Equation (3) is the difference-in-differences estimate obtained by DM after netting out confounding trends using the comparison areas. From the article, we know that the estimated difference (in-differences) in the mortality rate of New Orleans residents is between -0.5 and -0.2 (see Table 2 of text).

To identify the effect of Hurricane Katrina on the mortality of movers and stayers, it is necessary to know three values: $\alpha, (M_{post}^{stayers} - M_{pre}^{stayers})$, and $(M_{post}^{movers} - M_{pre}^{movers})$. The article provides estimates of α, but not the two other quantities. In 2006, approximately 50% of elderly and disabled residents of New Orleans had moved, but by 2013 approximately half had returned. Given these figures, I assume that the share of stayers (α) was 0.63 and the share of movers (1-α) was 0.37. Next, I calculate different values for $(M_{post}^{stayers} - M_{pre}^{stayers})$ conditional on assumed values for $(M_{post}^{movers} - M_{pre}^{movers})$. For example, evidence presented in the article suggests that the mortality rate of movers converged to the mortality rate of the place they moved to. The mortality rate of the areas that New Orleans residents moved to was approximately between 5.0 and 6.0 in 2006 (Table A.17 in text). Given these figures, a very large, perhaps implausibly large, estimate of $(M_{post}^{movers} - M_{pre}^{movers})$ is negative one percentage point. Using this estimate, the assumption that α=0.63, and the estimate of $M_{post} - M_{pre}$ (equation 3) provided by DM of -0.5, I derive an estimate of $(M_{post}^{stayers} - M_{pre}^{stayers})$ that is equal to -0.21. This calculation suggests that the mortality rate of stayers *decreased* by 0.21 percentage points.

Table 1 provides estimates of $(M_{post}^{stayers} - M_{pre}^{stayers})$ for different values of $(M_{post}^{movers} - M_{pre}^{movers})$ and for three estimates of $M_{post} - M_{pre}$ provided by DM that bracket other estimates. Under the assumption that Hurricane Katrina decreased mortality of movers ($M_{post}^{movers} - M_{pre}^{movers}$), as DM conclude, estimates in Table 1 strongly suggest that the mortality rate of stayers also declined in New Orleans after Hurricane Katrina, and possibly by more than the decline in mortality among movers. However, all evidence of the devastation and disruption caused by Hurricane Katrina suggest that the mortality of stayers should, if anything, increase.[2]

Table 1. Estimates of the Effect of Hurricane Katrina on Mortality of Movers and Stayers

| Effect on Movers (Assumed) $M_{post}^{movers} - M_{pre}^{movers}$ | Total Effect (Estimated by DM) $M_{post} - M_{pre}$ | Effect on Stayers (Derived) $M_{post}^{stayers} - M_{pre}^{stayers}$ |
|---|---|---|
| -1 | -0.5 | -0.21 |
| -0.8 | -0.5 | -0.32 |
| -0.6 | -0.5 | -0.44 |
| -0.4 | -0.5 | -0.56 |
| -0.2 | -0.5 | -0.68 |
| 0 | -0.5 | -0.79 |
| -1 | -0.35 | 0.03 |
| -0.8 | -0.35 | -0.09 |
| -0.6 | -0.35 | -0.20 |
| -0.4 | -0.35 | --0.32 |
| -0.2 | -0.35 | -0.44 |
| 0 | -0.35 | -0.56 |
| -1 | -0.2 | 0.27 |
| -0.8 | -0.2 | 0.15 |
| -0.6 | -0.2 | 0.03 |
| -0.4 | -0.2 | -0.08 |
| -0.2 | -0.2 | -0.2 |
| 0 | -0.2 | -0.32 |

---

[2] Appendix Figure A.13 in DM shows that mortality of stayers declined in the few years after Hurricane Katrina, although estimates were not statistically significant. However, this evidence comes from a problematic analysis that selects on moving status, which is endogenous. Stayers in New Orleans are unlikely to be comparable to stayers in other areas and DM provide no evidence to justify such an analysis.

What does the evidence in Table 1 suggest about the effect of place on mortality? Not much. Whether you stayed in New Orleans or moved to another place, mortality seemed to decline. The surprising finding is that mortality declined at all for either group. It surprising given the trauma (e.g., Laditka et al. 2010; Rhodes et al. 2010; Fussel and Lowe 2014; Calvo et al. 2015), widespread destruction of the healthcare infrastructure (Rittenhouse et al. 2012; Cole et al. 2015; Hamel et al. 2015) and destruction of the physical infrastructure and displacement that took place in varying degrees for both movers and stayers.

And there is other evidence casting doubt on the claim that the decline in mortality after Hurricane Katrina was due to moving. Estimates in Appendix Table A.7. show that movers were much more likely to be black (22 percentage points) and under age 65 (12 percentage points) than non-movers. If moving was the cause of the decline in mortality, then we would expect the mortality rates of black and under age 65 residents of New Orleans to go down more than that of other groups. Such is not the case. While the mortality rate of all black residents declined more than other race/ethnic groups after Katrina, the larger decline was not statistically significant. The mortality rate of the under age 65 residents of New Orleans was actually higher after Katrina than older groups, although this estimate was not significant. In general, there is little systematic relationship between the proportion of a demographic group that moved and the effect of Hurricane Katrina on mortality. This is further evidence suggesting that moving was not the primary cause of the decline in mortality reported.[3]

**Econometric Problems**

An important statistical issue that noticeably affects DM's results has to do with the way the regression model was specified. The issue is whether the model allows mortality rates to differ granularly by *age-by-year* as opposed to more coarsely by *age and year*, with the year effect being *the same* for all ages that year. The subtle distinction is important because the sample consists of a cohort of persons of different ages followed over time—meaning as they age. Older members of the cohort will die sooner than younger members and therefore the analysis should allow this to be the case. In most analyses, however, DM do not allow

---

[3] The effect of Hurricane Katrina on mortality depends on exposure (moving) and the effect of exposure, which may differ by demographic group. Thus, there does not have to be a correlation between the extent of exposure and the effect of Hurricane Katrina.

mortality rates to differ by age-by-year. When they do, results differ markedly.

*Model Specification*

The standard demographic model used to estimate time to death and applied to the DM context is:

$$DEATH_{jat} = \alpha_{at} + \sum_{t=2005}^{2013} \gamma_t (NO_j * YEAR_t) + \delta_j + e_{jat}$$

(4) $j = 1,...,J$ (ZIP codes)
$a = 30,...,85$ (age)
t=2004,...,2013

Equation (4) specifies that the probability of person of age "a" in 2004 in ZIP code "j" in year "t" dies (*DEATH*), conditional on being alive at the beginning of year t, depends on age-by-year fixed effects ($\alpha_{at}$), the interaction between a New Orleans dummy variable (*NO*) and year dummy variables, and ZIP code fixed effects.[4] An important feature of equation (4) is that there are separate year effects for persons of each age (here I use ages 30 to 85 because the youngest and oldest ages in the sample are not provided). The need to include these fixed effects stems from the fact that the probability that a person age 40 in 2004 is alive in any year after 2004 is very different than the probability a person 70 in 2004 is alive in any year thereafter. The older person has a much higher probability of dying in any year thereafter. The inclusion of age-by-year fixed effects is clearly the appropriate model, and it is the workhorse model of demographic studies of mortality.[5]

DM do not estimate this standard model for most analyses, but only in sensitivity analyses. Instead, DM estimate:

(1) $DEATH_{jat} = \alpha_a + \sum_{t=2005}^{2013} \gamma_t (NO * YEAR_t) + \delta_j + e_{jat}$

The key difference between equations (4) and (5) is that equation (5) restricts the year effects to be the same for each age.

---

[4] There is no justification for including ZIP code fixed effects. While they likely reflect differences in individuals and environment, they are arguably unnecessary to include given the inclusion of the New Orleans dummy variable.
[5] See for example, the Lecture notes of German Rodriguez found at: https://data.princeton.edu/wws509/notes/c7s6

As shown in DM's Appendix Figure A.7., the addition of age-race-sex-year fixed effects to the regression model results in substantially smaller estimates of the effect of Hurricane Katrina on mortality. For the 1992 and 1999 samples, the addition of such controls yields estimates that are not statistically significant in the period from 2009 (2010) to 2013. Thus, in these samples, any evidence of a decline in mortality post-Katrina may be due to what demographers and epidemiologists refer to as harvesting—the trauma of Hurricane Katrina speeded up the death (i.e., harvested) of some frail people and the spike in deaths after Katrina is naturally (biologically) followed by a temporary decline in mortality followed by a return to baseline. This is exactly what is shown in the plots in Figure A.1 of DM.

Perhaps more worrisome is that estimates from the standard and more appropriate demographic model reveal evidence of differential pre-trends—a sign of an invalid research design—that are non-trivial in magnitude when compared to the post-Katrina estimates.

As noted, estimates reported by DM from equation (4) are substantially smaller and often not statistically significant relative to estimates from equation (5). Why would that be the case? Consider an extreme case in which all persons in New Orleans are age 40 and all persons in comparison areas are age 60. For any year post-Katrina the probability of dying would be lower for the younger residents of New Orleans than the older residents of other places even conditioning on the baseline age difference. This is consistent with the different estimates from equations (4) and (5) reported by DM and the fact that residents in New Orleans are younger than residents in other areas. Differences in race and gender may have similar confounding influences. The upshot is that when the correct model is used, estimates of the effect of Hurricane Katrina on mortality are much smaller (half the size or less) and less statistically significant. Such findings suggest the presence of the harvesting effect.

*Selective Mortality and Unmeasured Heterogeneity*

A notable empirical fact reported by DM is that *all* residents in New Orleans had higher rates of mortality than most of the comparison areas (Figures 1 and A.5. in text). Here I return to the first of my initial illustrations. Consider two individuals who are both age 70 (approximately the mean age of sample) in 2004, but one lives in New Orleans and the other lives in Richmond. The higher rate of mortality in New Orleans suggests that the 70-year-old in New Orleans is healthier than the 70-year-old in Richmond because of selective mortality—to make it to age 70 in New Orleans with its relative mortality rate a person has to be healthier than a 70-year-old in the lower-mortality area.

Evidence that selective mortality is an issue is found in DM's Appendix Figure A.7., which shows diverging pre-trends between the mortality rate in New Orleans and other areas, with New Orleans having, as people age, a consistently larger decline in mortality than the comparison areas prior to Hurricane Katrina. The magnitudes of the deviations between New Orleans mortality and mortality in other areas prior to Katrina are about half as large as the post-Katrina deviations. Again, we find substantial evidence to question the basic claim that Hurricane Katrina reduced mortality.

*Inference*

DM's approach to inference is questionable. The approach is the so-called cluster-robust method. DM construct standard errors allowing for non-independence of observations within baseline ZIP code without providing any justification for their choice. Moreover, it is not a statistically sound choice and there are available alternatives. For analyses that use cohorts from 1992 or 1999 and the 10-city comparison group, the approach of Donald and Lang (2007) would be feasible and preferable. That approach was not used, however, and DM do not explain why it was not used. For analyses that use counties as comparisons for New Orleans (N=152), or Commuting Zones (N=400), instead of the 10 comparison cities, the method of Conley and Taber (2011), or randomization inference (e.g., Good 2005) could be used, but was not.

It is well known that standard errors can be substantially different. To see the potential problem, consider DM's main analyses using the 2004 cohort and residents of New Orleans and 10 comparison cities. One view of these data is that there are only six "observations"—three observations for New Orleans for 2004, 2005 and 2006-2013 and three observations for the comparison cities in 2004, 2005, 2006-2013 (as in Table 2). Table 2 in DM reports sample sizes of roughly 8 and 10 million observations, which are not very relevant, and notes that standard errors are clustered by ZIP codes with no justification. One may argue that inference should be conducted assuming six observations, which is to say that no inference can be made (Donald and Lang 2007; Cameron and Miller 2015).

Instead of applying the just-noted two feasible approaches to inference to the difference-in-differences analyses that produced the estimates underlying the article's conclusions, DM chose to alter the research design, apparently just for the purpose of conducting inference. It is a curious choice. Specifically, DM use a synthetic control (SC) approach instead of a difference-in-difference approach to estimate the effect of Hurricane Katrina on mortality. While the inference from the SC approach suggests that Hurricane Katrina had a significant, negative effect on mortality, the magnitudes and pattern of estimates differs markedly from those obtained from the difference-in-differences approach. Estimates are much larger (nearly twice as large) and estimates do not become smaller over time, as do the difference-in-differences estimates (from the best practice specification).

Applying the conclusion about statistical significance based on the inference approach from this supplemental analysis, which did not undergo any sensitivity analysis (e.g., in the method to select weights) and that is only cursorily described, to estimates obtained from the difference-in-differences analysis seems a bridge to far. The authors offer no justification for applying the SC-derived conclusion to the difference-in-difference analysis. More importantly, there is no reason that randomization (permutation) inference (or the approach of Conley and Taber 2011) could not have been conducted in the context of a difference-in-differences analysis—there was no need to move to a different research design to conduct such inference (e.g., Kaestner 2016). Finally, no SC analysis was conducted for the 2004 cohort because it is infeasible, but the alternative approaches to inference could have been used for this cohort. In sum, the lack of a valid approach to inference means that none of the primary estimates reported by DM can be assessed from a statistical point of view. They do not provide reliable evidence of the effect of Hurricane Katrina on mortality.

**No Theory-No Hypotheses? No Problem—We Have a Research Design**

The DM article is almost entirely an empirical exercise with virtually no attempt to use theory to generate testable hypotheses. The only theoretical statement I could find is:

> The disruption induced by extreme weather events can be used to illuminate factors that affect the accumulation or depreciation of health capital (Grossman1972). (DM, 3602-03)

Without any theorizing about how purposive human beings would bring about putatively observed results, the plausibility of estimates depends solely on the credibility of the research design. The interpretation of estimates becomes ad hoc.

DM refer to the finding that Hurricane Katrina reduced mortality of New Orleans residents as counterintuitive. The finding is more than counterintuitive. It is inconsistent with the canonical model of the demand for health (Grossman 1972). The Grossman (1972) model would predict an increase in mortality for residents that remained in New Orleans because of:

- Worsened health after Katrina because of increased stress and adverse environmental factors such as mold (Laditka et al. 2010; Rhodes et al. 2010; Fussel and Lowe 2014; Calvo et al. 2015).
- A decrease in the productivity of investment in health and degradation of the health care system (Rittenhouse et al. 2012; Cole et al. 2015; Hamel et al. 2015).
- Higher costs to investing in health, for example, because of less access and availability to medical services and higher prices of goods and services (e.g., rents rose significantly[6]).
- A decline in income because of loss of employment and macroeconomic effects.[7]

As shown in Table 1, estimates and other information found in DM suggest that mortality declined among residents who remained in New Orleans. How should we interpret this? Is it a surprising, novel finding that should make us rethink theory? Or should we question the credibility of the estimate? The atheoretical analysis and legitimate questions about the validity and reliability (inference) of the estimates in DM leads me to conclude that we should doubt the estimate.

---

[6] See: https://financialservices.house.gov/uploadedfiles/fact_sheet_status_of_housing_in_nola_10_yrs_later_final.pdf

[7] The analysis of the effect of Hurricane Katrina on labor income of New Orleans residents in Deryugina et al. (2018) shows that over the period from 2005-13 labor income was statistically no different from the 2004 level. The method of inference used in this study suffers from the same problems as described earlier. Labor income is not as important for elderly and disabled because many do not work. Net income likely declined for these demographic groups because of increased expenditures.

Theory also can be used to assess whether the mortality of those who left New Orleans is likely to decrease. The Grossman (1972) model suggests the mortality of movers will be affected by worsened health, the access and price of health care, the quality of health care, and income.

Consider worsened health. It is unlikely that the age pattern of onset of illness changes much with destination location, although pollution and other environmental factors may be present. Interestingly, DM show that moving to a low-pollution area is associated with an *increase* in mortality (Figure 6). DM also report that movers are differentially selected with respect to destination pollution—healthier movers are significantly more likely to go to more polluted places. Movers are also significantly selected on upward income mobility and median house value. Several other characteristics (social capital, crime rate, income segregation, and urban population) show non-trivial though only marginally significant correlations between mover characteristics and destination characteristics. Of course, economic theory suggests that a mover's destination is not random even if the decision to move were random.

To support the argument that moving is random (and to overturn standard economic theory), DM rely on the fact that mover's predicted mortality based on race (black), sex, age (<64, 75+) and several chronic condition indicators is not strongly correlated with destination mortality. However, these characteristics likely explain a small portion of mover mortality (not reported by DM) and this analysis is not very strong evidence against the possible bias due to the likely non-random sorting of mover's by destination, particularly because there is substantial evidence that mover's destination was non-random. Others have provided similar evidence. Callison et al. (2020), for example, show that among elderly Medicare enrollees, sicker movers are more likely than healthier movers to select high-spending Medicare destinations.

The findings with respect to pollution are of particular importance because it is one of the few destination characteristics with a clear causal link to mortality. Most other destination characteristics have no such link. Consider the health behaviors of movers now residing in new destinations.

Estimates in Figure 6 of DM show that moving to a high-smoking area, or a high-obesity area, increases mortality, and moving to a high-exercise area decreases mortality. The causal mechanisms underlying these associations is unclear. Few people actually quit smoking. DM provide no evidence that the level of smoking in an area causes someone (e.g., a mover) to smoke less. Indeed, the trauma and stress associated with Katrina might lead one to expect an increase in smoking among movers. Even if there were differences in the cigarette prices and tobacco control policies in an area, which DM could have, but did not measure, evidence suggests that these policies would have very minor effects on elderly smoking (e.g., Callison and Kaestner 2014). Similarly, why would area obesity level cause a mover's mortality to increase? Does a 70-year-old suddenly lose weight because there is less obesity in the area she moved to? That is doubtful, and DM provide no evidence of such behavioral channel.[8] Does an elderly or disabled person start jogging because they see some people jogging near their house? Again, this is doubtful. The lack of strong causal links between the characteristics that were considered and mortality is obvious and reveal the lack of a theoretical basis of the mover analysis and the ad-hoc, empirical approach taken. Imbuing estimates of these correlations with causal meaning is inappropriate and likely misleading.

Consider the access and price of medical care. All persons in the sample are covered by Medicare, and a much higher proportion of New Orleans residents are covered by Medicaid as dual eligible and therefore have more generous benefits. This fact implies that there are virtually no price differences or access differences between New Orleans and destination areas that would significantly affect mover's mortality. Does the quality of care differ markedly between New Orleans and destination areas? Perhaps, but this is unmeasured. Here too, estimates in DM's Figure 6 suggest the answer is no, as measures of the health care infrastructure (not necessarily quality) in destination places is unrelated to mover mortality.

---

[8] Datar and Nicosia (2018) examine associations between the county obesity rate and weight status of enlisted (military) adults (mean age 37) who are arguably randomly assigned to areas. Associations are positive, but small and usually not statistically significant. Similarly, Christakis and Fowler (2007) and Christakis and Fowler (2008) report that the obesity and smoking of neighbors is not related to either of these health behaviors.

Finally, consider income. Yet again, estimates in DM's Figure 6 show that moving to an area with a higher per-capita income decreases mortality, but moving to an area with high rates of elderly poverty has no effect. But most movers were not working, so, here too, the causal mechanisms between area income and mover mortality are lacking.

Overall, there is little evidence consistent with theory to support the conclusion that moving reduced mortality. So, why do DM conclude that moving reduced mortality? The main reason underlying the claim is the estimate of the association between destination mortality and the probability of dying. But why should we trust this estimate in light of the documented non-random selection of mover's to destinations, the inconsistent evidence (pollution and social capital) and the fact that there is no theory underlying this particular association. Why should destination mortality affect mover mortality? What is the causal mechanism? None is provided, and when theoretical causes of mortality are assessed, the evidence is weak or inconsistent with theory. In short, the empirical analysis conducted by DM on movers is, at best, exploratory, and at worst, an atheoretical, data mining exercise.

**Conclusion**

The surprising finding reported in DM that a hurricane as devastating as Katrina reduced the mortality of residents of New Orleans merits close scrutiny because it is inconsistent with intuition, theory, and prior evidence (e.g., Laditka et al. 2010; Rhodes et al. 2010; Fussel and Lowe 2014; Calvo et al. 2015). I have provided a thorough review and critical assessment of the evidence provided by DM to support their conclusions. I find the evidence wanting.

First, I show that based on the estimates reported in DM and reasonable assumptions, DM's evidence would suggest that the mortality rate of residents of New Orleans *who remained in New Orleans* decreased after Katrina. While possible, there is no external evidence or theory to support this conclusion. Moreover, when the appropriate regression model is used (equation 4), estimates of the effect of Hurricane Katrina on mortality are greatly reduced in magnitude and there is evidence of diverging pre-trends that raise questions of the validity of the difference-in-differences research design. Then there is the serious problem related to how DM approached inference. They simply never provided a valid approach to inference despite the availability of such approaches widely utilized in the literature.

Second, an assessment of the mover analysis and the effects of destination characteristics on mover mortality reveals non-trivial selection and several inconsistencies. Moving to a place with lower pollution is associated with an increase in mortality. Note that pollution is the only direct environmental influence assessed by DM. This is important because there are, at best, distal causal links between area characteristic such as obesity, smoking, exercise and income and a mover's mortality. Thus, the conclusion by DM that moving reduced mortality, which is something never actually assessed directly, is unjustified.

I will end by questioning what we expect to learn from an atheoretical, empirical exercise of a one-off natural disaster. There will never be another Hurricane Katrina in New Orleans. New Orleans has already changed physically and institutionally, and the composition of the city's population has changed. Thus, even if another storm of the same magnitude rolled through New Orleans, it is unlikely that findings from this study would apply. The external validity of the findings, even assuming that the findings are valid, to other cities and other storms seems extremely limited. Cities differ in their infrastructure, institutions, and populations (see Table 1 in DM which show substantial differences even among purposely matched cities). These and many other factors determine the effects of a natural disaster, such as a hurricane. As Nancy Cartwright (2013) argues, in most cases evidence from one experiment, in this case a natural experiment, do not travel well and are unlikely to be replicated in another (natural) experiment. The near absence of any plausible external validity and the absence of any tests of theoretically derived hypotheses make studies like DM's marginal in their scientific importance.

DM ask in their title: Does when you die depend on where you live? Surely it does, in some cases, and we do not need a study of hurricane related trauma and displacement from New Orleans to know that. It is well documented that there are environmental causes of mortality, such as pollution, unsafe drinking water, poor sanitation, disease (e.g., malaria), and crime. It is well known that environments differ geographically. Similarly, there are longstanding geographic differences in social and economic opportunities that vary by geography (e.g., Appalachia) that are a cause of low-incomes, low-education, and poor health behaviors. The geographical landscape of disease and mortality in the U.S. has been thoroughly assessed by the Centers for Disease Control since at least 1975 (Mason et al. 1975; Mason et al. 1976; Mason et al. 1981; Pickle et al. 1990; Pickle et al. 1996). So, the interesting question is not whether where you live affects when you die, but how to affect the causes of mortality and how to facilitate mobility to diminish the contribution of geography-related mortality. The DM study is silent on these questions.